\documentclass[twocolumn,groupedaddress,floatfix,pra,showpacs]{revtex4}
\usepackage{amsmath,amssymb ,bm}
\usepackage{graphicx}
\usepackage{epstopdf}
\usepackage{float,braket}
\usepackage{color,dsfont}
\usepackage{appendix}
\usepackage{lipsum}
\usepackage[T1]{fontenc}
\usepackage[utf8]{inputenc}
\usepackage{lmodern,hyperref}
\usepackage{graphicx,amsmath,amsthm,amssymb,epstopdf,color,bm}

\begin{document}

\title{Practical Quantum Teleportation of an Unknown\\ Quantum State}

\author{Syed Tahir Amin}
\author{Aeysha Khalique}
\affiliation{%
School of Natural Sciences, National University of Sciences and Technology,
	H-12 Islamabad, Pakistan
	}
\pacs{03.67.-a, 03.67.Hk, 03.67.Bg}

\date{\today}

\begin{abstract}
We develop a theory to teleport an unknown quantum state using entanglement between two distant parties. Our theory takes into account experimental limitations due to contribution of multi-photon pair production of parametric down conversion source, inefficiency and dark counts of detectors and channel losses. We use a linear optics setup for quantum teleportation of an unknown quantum state by performing Bell state measurement by the sender. Our theory successfully provides a model for experimentalists to optimize the fidelity by adjusting the experimental parameters. We apply our model to a recent experiment on quantum teleportation and the results obtained by our model are in good agreement with the experiment results.

\end{abstract}

\maketitle

\section{Introduction}
\label{sec:intro}

Quantum teleportation~(QT) is the process of sending quantum state of any physical quantum
system from one place to another place without sending the system itself. The idea of~QT, presented in~1992~\cite{BBCJ+93} consisted of disassembling of an unknown quantum state into purely classical information and quantum correlations and then later reconstructing the same quantum state by using this information. 

Quantum teleportation is a key tool in quantum computation~\cite{GC99}. Many successful experimental attempts have been made to demonstrate~QT, after the first demonstration in~1997 using entangled photons~\cite{BPME+97}. Laboratory demonstrations involve open destination~QT~\cite{ZCZY+04}, entanglement swapping demonstration~\cite{PBWZ98}
and two-bit composite system~QT~\cite{ZGWC+06}. In addition QT through fiber link has been realized~\cite{MRTZ+03,UJAK+04} and is limited to~1~km.
Recently,~QT over 16~km is demonstrated via free space links~\cite{JRYY+10} using
single entangled photons pair and over~100~km using parametric down conversion sources. Although~QT demonstrated by above techniques transmits a state having an overlap with original state above the classical limit, but still these~QT techniques are unable to proceed with 100\% overlap. It is thus important to see the effect of various experimental parameters limiting the~QT process.

Quantum teleportation has been analyzed before for parametric down conversion~(PDC) sources for entangled photon pairs but with approximate sources~\cite{KB00}. In this paper we thoroughly analyze the multipair effect of the~PDC sources including detector dark counts and inefficiencies. We have also considered the limiting factors in long distance~QT. Our theory relies on the model of resources presented in~\cite{SHS+09}, where Bayesian approach is used to relate the counts on ideal detectors with unit efficiency and zero dark counts to inefficient detectors with dark counts. Our results show that multipair production is counter-productive above certain value, though a high fidelity can be achieved for any value of detector inefficiency for very low dark counts and low pair production rate. This pair production rate is much lower than the one practically achievable. Our model fits well to a recent experiment~\cite{YRLC+12}, which has achieved teleportation over a distance of~100~km. This allows us to analyze and compare our model with real world experiments for long distance~QT. Our model is useful to optimize various experimental parameters for maximizing fidelity for~QT to any distance.

Our paper proceeds as follows: In Sec.~\ref{sec:background},  we provide
background on practical resources, which concerns about sources generating entanglement and detectors models. In Sec.~\ref{sec:theoryofPQT}, we compute the teleported quantum state using threshold detectors  and incorporating inefficient entangled sources by Bell state measurement. In Sec.~\ref{comparison}, we apply our model to a recent experiment and show the agreement between the two. Finally, we summarize our results and conclude in Sec.~\ref{sec:conclusions}.

\section{Background: QT and practical resources}\label{sec:background}
Entanglement is a resource in quantum communication~(QC). Communication protocols like quantum teleportation rely on entanglement for long distance distribution of an unknown quantum state. The entanglement generation sources, however produce multiple pairs which effect the distributed state. Other affecting factors are the inefficient detectors with dark counts. In this section we prepare a background for our model of QT including the above mentioned resources imperfections. We give a brief introduction to quantum teleportation process in Sec.~\ref{sec:techqt} and  review the model of the resources in Sec.~\ref{sec:resources}.

\subsection{Technique of QT}
\label{sec:techqt}
\begin{figure}[h]
\includegraphics[width=8cm,height=5.1cm]{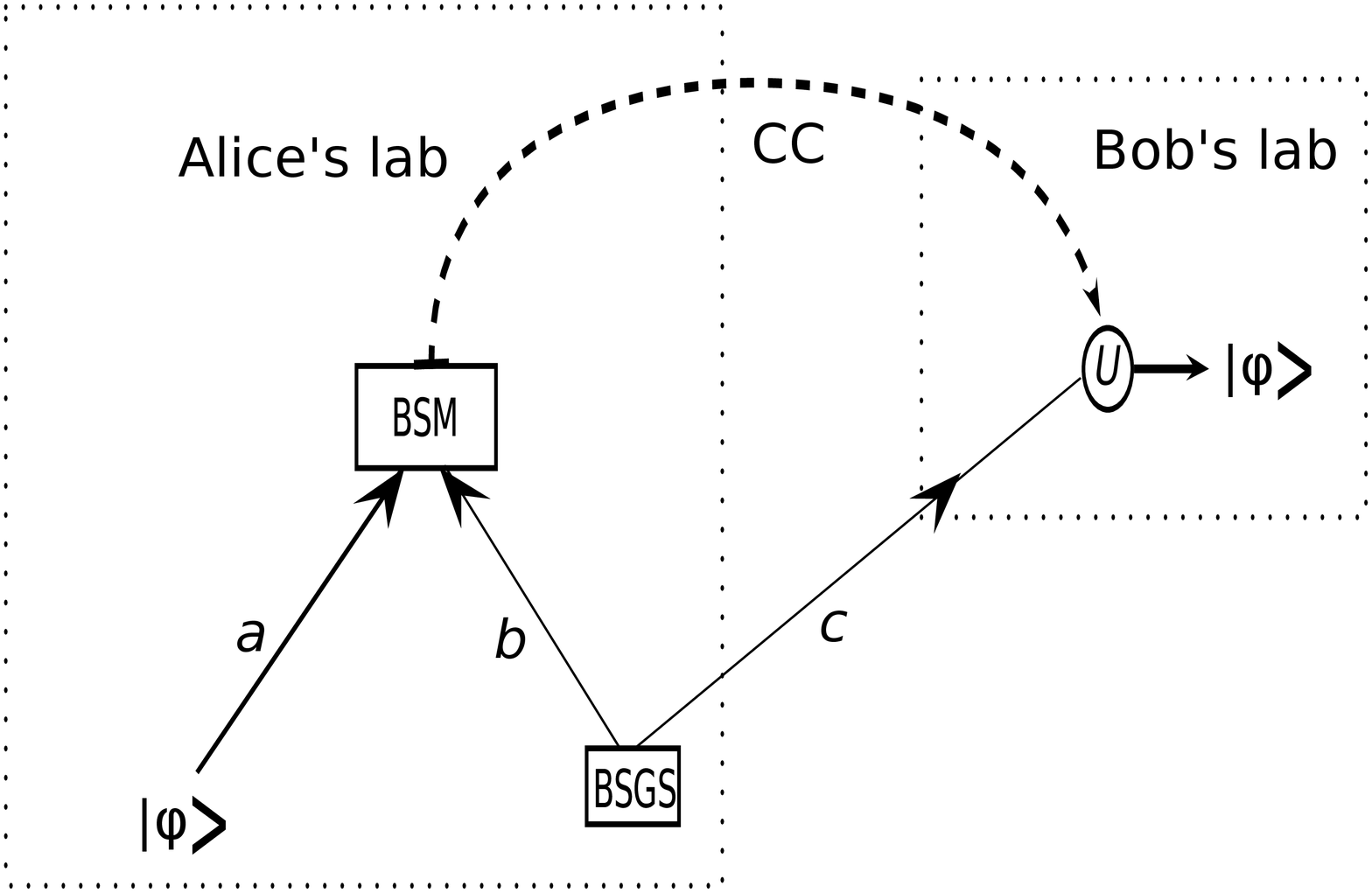}
\caption{ Quantum teleportation:  Bell State Generation Source (BSGS) produces an entangled pair in one of the Bell states in modes $b$ and $c$. Unknown state to be teleported, $\ket{\varphi}$ at mode $a$ is combined at Bell State Measurement (BSM). The result is communicated to Bob through classical communication (CC), who performs an appropriate unitary $U$ to get $\ket{\varphi}$
}
\label{fig:qt}
\end{figure}

Quantum teleportation is sending of quantum state of any physical system from one place to another without sending physical system itself. Sender, Alice, wants to send an unknown quantum state of a photon, $\ket{\varphi}=\alpha|0\rangle +\beta|1\rangle$, to receiver Bob, where both of them are spatially separated. It is not possible to know the state of the photon by direct measurement and quantum no-cloning prohibits making copies of the state. Transmission of the state relies on the idea given in 1992~\cite{BBCJ+93}, where Alice and Bob use an entangled pair together with the unknown state. The process of teleportation is depicted in Fig.~\ref{fig:qt}.
\begin{equation}
\ket{\phi^{\pm}}=\frac{1}{\sqrt{2}}\left(\ket{00}\pm\ket{11}\right)\quad\ket{\psi^{\pm}}=\frac{1}{\sqrt{2}}\left(\ket{01}\pm\ket{10}\right),
\end{equation}
which are the maximally entangled states. Alice then performs
a joint measurement, the Bell state measurement on the unknown quantum system and on
her part of the shared entangled pair, which collapses in one of the above mentioned Bell states. She then sends the result to Bob through classical communication. For the entangled pair in Bell state $\psi^-$, Bell state measurement resulting in $\psi^-$, projects Bob's part of the initially shared pair in the same state that Alice wanted to send. 
For Bell measurement resulting in $\phi^-$, $\phi^+$ and $\psi^+$ at Alice, Bob performs the unitaries ($U$), $\hat\sigma_x$, $\hat\sigma_y$ and $\hat\sigma_z$, respectively, on his particle, where
\begin{align}
\sigma_x&=\ket{0}\bra{1}+\ket{1}\bra{0},\nonumber\\
\sigma_y&=-i\ket{0}\bra{1}+i\ket{1}\bra{0},\nonumber\\
\sigma_z&=-\ket{1}\bra{1}+\ket{0}\bra{0},
\end{align}
are the Pauli operators, which transform Bob's particle in the same state as the desired unknown state. However, because of the imperfect resources teleportation is not perfect. We discuss these imperfect resources in next section.

\subsection{Modeling of resources used in QT}
\label{sec:resources}
In this section, we give a review of model of various resources, used in QT, including imperfections. 

Ideally single pair sources should be used as entangled pair source, but they are not available, so practically spontaneous parametric down conversion (SPDC) sources are used. Here we give a model of practical SPDC TYPE-II source. Such a source emits polarization entangled photon pairs in some spatial modes, which are labeled as \textquoteleft $b$\textquoteright\space and \textquoteleft $c$\textquoteright\space. These photons have orthogonal polarization labeled as horizontal~($H$) and vertical~($V$). Basically we want to find out state generated by SPDC TYPE-II, which is similar to the  following  Bell state, 
 \begin{align}
   |\psi^{-}\rangle_{bc}=&\frac{1}{\sqrt{2}}(|HV\rangle-|VH\rangle)_{bc},\nonumber\\
   =&\frac{1}{\sqrt{2}}(|1001\rangle-|0110\rangle)_{b_{H}b_{V}c_{H}c_{V}}.
 \end{align}  
 The complete quantum state prepared by such SPDC TYPE-II~\cite{BRSJ+01}  is given as
 \begin{align}
   |\chi \rangle\label{eq:pdcunnormalize}=&\exp[i\chi(b_{H}^{\dag}c_{V}^{\dag}-b_{V}^{\dag}c_{H}^{\dag}+b_{H}c_{V}-b_{V}c_{H})]|\hbox{vac}\rangle,
 \end{align}
 where, $\chi^2$ is proportional to the efficiency of generation of entangled photon pairs or photon pair production rate of the source. Normal order form of above equation is~\cite{SHS+09},
 \begin{align}
 |\chi\rangle=&\label{eq:PDC}\exp[2\omega(\chi)]\exp[\phi(\chi)(b_{H}^{\dag }c_{V}^{\dag }-b_{V}^{\dag }c_{H}^{\dag })]|\hbox{vac}\rangle.  
 \end{align}
Note that $|\hbox{vac}\rangle=|0_{b_{H}}0_{b_{V}}0_{c_{H}}0_{c_{V}}\rangle$.
Ideally quantum state of modes $b$ and $c$ correspond to an entangled state of photons pair but in realistic scenario it contains vacuum state and higher order Fock states as well. 

Here we reprise the mathematical formalism of detecting $q$ photons such that $i$ photons are incident onto threshold detector, which click if there are photons in certain mode and do not click if there are no photons in that mode. Such a detector with efficiency $\eta$ and dark counts probability $\zeta_{dc}$, does not click with probability
\begin{align}
  p_{\zeta _{dc},\eta}\label{eq:pbofnodetection}(\hbox{no click}|i)&=p_{\zeta _{dc},\eta}(q=0|i)\nonumber\\&=(1-\zeta _{dc})\{1-\eta(1-\zeta _{dc})\}^{i}.
\end{align}
The first term in Eq.~\eqref{eq:pbofnodetection} is the probability that threshold detector does not click and there is no dark count, and the second term is the probability that $i$ photons are incident on detector but detector does not detect any incident photon. The probability of getting a click is then 
\begin{align}
   p_{\zeta _{dc},\eta}\label{eq:pbofdetection}(\hbox{click}|i)&=1-p_{\zeta _{dc},\eta}(\hbox{no click}|i)\nonumber\\&=1-(1-\zeta _{dc})\{1-\eta(1-\zeta _{dc})\}^{i}.
\end{align}

There are four detectors, one for each of the four spatial modes. Using threshold detectors, the posterior conditional probability for any read out $(ijkl)$, which four ideal detectors, with unit efficiency and zero dark counts, would have yielded is~\cite{SHS+09}
  \begin{align}\label{eq:Bayes}
P_{ijkl}^{qrst} &=P (ijkl|qrst) \nonumber\\&=\frac{p(qrst|ijkl)p(ijkl)}{\sum^{\infty }_{i_{1}j_{1}k_{1}l_{1}}p(qrst|i_{1}j_{1}k_{1}l_{1})p(i_{1}j_{1}k_{1}l_{1})}.
 \end{align}
 Here, $p(ijkl)$ is the probability that ideally $\{ijkl\}$ photons are measured on the four detectors. As the four detectors are independent of each other, therefore
  \begin{align}\label{eq:independentprobabilities}
    p(qrst|ikjl)&=p(q|i)p(r|j)p(s|k)p(t|l).
  \end{align}  
 All transmission and other coupling losses are included in detector efficiency.
 
 Equipped with the model of the resources, we present our model of quantum teleportation in next section.
 
\section{Modeling of  practical QT}\label{sec:theoryofPQT}
 We now develop our model of~QT under practical conditions of the resources discussed in last section. In Sec. \ref{postmeasuredSID} we compute the pure teleported quantum state and its conditioned probability using ideal photon discriminating detectors in Bell state measurement. In Sec.~\ref{postmeasuredSTD}, we relate this state to that produced by practical threshold detectors in the form of mixed density matrix. We then find out the fidelity of teleported state in Sec.~\ref{Fidelity}.

\subsection{Postmeasured teleported quantum state using ideal detectors}
\label{postmeasuredSID}
Using the resources presented in the previous section we are
now in a position to derive the teleported quantum state. We start with quantum state produced by PDC-Type II source
given in Eq.~\eqref{eq:PDC} and unknown quantum state in spatial mode $a$ as $|\varphi\rangle_{a}=\alpha|0\rangle + \beta|1\rangle$, which in $H-V$ modes takes the form $|\varphi\rangle_{a}=\alpha|1_H 0_V\rangle + \beta|0_H1_V\rangle$. The composite state of photons in modes $a$, $b$ and $c$ will be 
\begin{align}
  |\psi\rangle_{abc}=&|\varphi\rangle_{a}\otimes|\chi\rangle_{bc}\nonumber\\
  =&\alpha\exp[2\omega (\chi)]\exp[\phi(\chi) (b^{\dag }_{H}c^{\dag }_{V})]\nonumber\\&\times\exp[-\phi(\chi) (b^{\dag }_{V}c^{\dag }_{H})]a^{\dag }_{H}|\hbox{vac}\rangle_{abc}\nonumber\\&+\beta\exp[2\omega (\chi)]\exp[\phi(\chi) (b^{\dag }_{H}c^{\dag }_{V})]\nonumber\\&\times\exp[-\phi(\chi) (b^{\dag }_{V}c^{\dag }_{H})]a^{\dag }_{V}|\hbox{vac}\rangle_{abc}.
\end{align}
 Bell state measurement is then performed by combining modes $a$ and $b$ on a balanced beam splitter. The four tuple of detectors then measure the photons in the two polarization modes in each of $a$ and $b$. First we consider that the four detectors used in Bell state measurement were perfect, having unit efficiency~$(\eta=1)$ and no detector dark counts$~(\zeta_{dc}=0)$. Applying balanced beam splitter transformation $U_{{\tiny \hbox{BS}}}$ to modes $a$ and $b$ using the rule~\cite{SHS+09}
\begin{align}
  a^{\dag }_{H}\xrightarrow{U_{\tiny{\hbox{BS}}}} \frac{1}{\sqrt{2}}(a^{\dag }_{H}-b^{\dag }_{H}),&& b^{\dag }_{H}\xrightarrow{U_{\tiny{\hbox{BS}}}}\frac{1}{\sqrt{2}}(a^{\dag }_{H}+b^{\dag }_{H}),\vspace{-.9cm}\nonumber\\
  a^{\dag }_{V}\xrightarrow{U_{\tiny{\hbox{BS}}}}\frac{1}{\sqrt{2}}(a^{\dag }_{V}-b^{\dag }_{V}),& & b^{\dag }_{V}\xrightarrow{U_{\tiny{\hbox{BS}}}}\frac{1}{\sqrt{2}}(a^{\dag }_{V}+b^{\dag }_{V}).
 \end{align}
The resultant three spatial modes quantum state, $U_{\tiny{\hbox{BS}}}|\psi\rangle_{abc}$, after passing through balanced beam splitter is projected onto subspace using the projection operator 
\begin{align}\label{eq:projection}
  \Pi^{(ikjl)} _{a_{H}a_{V}b_{H}b_{V}}:=& (|i\rangle\langle i|)_{a_{H}}\otimes (|j\rangle\langle j|)_{a_{V}}\otimes (|k\rangle\langle k|)_{b_{H}}\nonumber\\&\otimes (|l\rangle\langle l|)_{b_{V}}\otimes{I} _{c_{H}} \otimes {I} _{c_{V}}.
\end{align}
Here $(|n\rangle\langle n|)_{d_{H}}$, represents projection operator on some Fock state $|n\rangle$ in mode $d$ having horizontally polarized photon and analogously for vertically polarized photon. The post measured quantum state obtained by state normalization after  projection operator~\eqref{eq:projection}  is
\begin{eqnarray}\label{eq:postmeasuredstate}
\frac{\Pi^{ijkl} _{a_{H}a_{V}b_{H}b_{V}}(\textbf{\textit{U}}^{ab}_{\tiny{\hbox{BS}}}|\psi \rangle_{abc})}{\sqrt{p(ijkl)}}=&|ijkl\rangle_{a_{H}a_{V}b_{H}b_{V}}\otimes|\Phi \rangle^{ijkl}_{c_{H},c_{V}},\nonumber\\
\end{eqnarray}
 with the first factor being the Fock state with modes $a_{H},\  a_{V},\ b_{H},\ b_{V}$ having  $i$, $j$, $k$, $l$ photons respectively, and
\begin{widetext}
\begin{align}
\label{eq:pureteleportedstate}
&|\Phi \rangle^{ijkl}_{c_{H},c_{V}} = \frac{1}{\textbf{[}\alpha^{2}(i+k-1)!(j+l)!(i-k)^{2}+\beta^{2}(i+k)!(j+l-1)!(l-j)^{2}\textbf{]}^{\frac{1}{2}}}\nonumber\\
&\times\textbf{[}\alpha\sqrt{(i+k-1)!(j+l)!}(i-k)|j+l,i+k-1\rangle+\beta\sqrt{(i+k)!(j+l-1)!}(l-j)|j+l-1,i+k\rangle\textbf{]}_{c_{H},c_{V}}.\nonumber\\
\end{align}
\end{widetext}
is the state on spatial mode $c$ such that $\{ijkl\}$ photons are detected on the other two modes.  Equation(\ref{eq:pureteleportedstate}) thus gives the teleported state.
The corresponding probability of the hypothetical ideal measurement readout $(ijkl)$
\begin{align}\label{eq:idealprobability}
 p(ijkl)&=\frac{[\tanh\chi]^{2(i+j+k+l-1)}}{\cosh ^{4}\chi (2^{i+j+k+l}i!j!k!l!)}\nonumber\\&\times\textbf{[}\alpha^{2}(i+k-1)!(j+l)!(i-k)^{2}\nonumber\\&+\beta^{2}(i+k)!(j+l-1)!(l-j)^{2}\textbf{]}.
\end{align}
is the probability that pure quantum state $|\Phi \rangle^{ijkl}_{c_{H},c_{V}}$ will be detected using ideal detectors in Bell state measurement. 
\subsection{ Postmeasured teleported quantum state using threshold detectors}
\label{postmeasuredSTD}
We find out teleported quantum state in previous section, using ideal detectors having efficiency $\eta=1$ and dark counts probability $\zeta_{dc}=0$. Since in practical scenario the detectors are not perfect but have efficiency, $\eta <1$, and non-zero dark counts, $\zeta_{dc}\ne 0$, we can neither find the post-measured pure resultant quantum state nor the probability of its occurrence of the remaining mode $c$. However for any read out $(ijkl)$ of four tuple detectors we can calculate its posterior probability $P^{qrst}_{ijkl}=p(ijkl|qrst)$ using Bayes theorem \eqref{eq:Bayes}. We can achieve our task using the conditional probabilities $p(qrst|ijkl)$ in~Eq.~\eqref{eq:independentprobabilities}. Hence, the resultant teleported quantum state of the remaining mode $c$, after Bell state measurement  using threshold detectors yielding actual read out $(qrst)$, is a mixed state of the form
\begin{align}
\label{eq:state}
 \varrho^{qrst}_{c_{H},c_{V}}=& \sum_{ijkl}P_{ijkl}^{qrst} |\Phi \rangle^{ijkl}_{c_{H},c_{V}}\langle\Phi|, 
\end{align} 
Using Eqs.~\eqref{eq:idealprobability}~and~\eqref{eq:Bayes}, we calculate
\begin{align}\label{eq:posteriorprobability}
    P_{ijkl}^{qrst}=&\frac{p(qrst|ijkl)[\tanh\chi]^{2(i+j+k+l)}}{Z^{\tiny{qrst}}\times(2^{i+j+k+l}i!j!k!l!)}\nonumber\\&\times\textbf{[}\alpha^{2}(i+k-1)!(j+l)!(i-k)^{2}\nonumber\\&+\beta^{2}(i+k)!(j+l-1)!(l-j)^{2}\textbf{]},
  \end{align}  
  with 
  \begin{align}\label{eq:fracprobability}
Z^{\tiny{qrst}}= \sum^{\infty}_{i_{1}j_{1}k_{1}l_{1}}&\frac{p(qrst|i_{1}j_{1}k_{1}l_{1})[\tanh\chi]^{2(i_{1}+j_{1}+k_{1}+l_{1})}}{2^{i_{1}+j_{1}+k_{1}+l_{1}}i_{1}!j_{1}!k_{1}!l_{1}!}\nonumber\\
&\times E^{(i_{1}j_{1}k_{1}l_{1})},
  \end{align}
where,
\begin{align}
    E^{(i_{1}j_{1}k_{1}l_{1})}=&\textbf{[}\alpha^{2}(i_{1}+k_{1}-1)!(j_{1}+l_{1})!(i_{1}-k_{1})^{2}\nonumber\\&+\beta^{2}(i_{1}+k_{1})!(j_{1}+l_{1}-1)!(l_{1}-j_{1})^{2}\textbf{]}.
\end{align}
Thus we have developed a closed form of the teleported state at Bob's end incorporating faulty detectors and multipair SPDC sources. This state is not exact replica of the desired state to be sent due to the imperfections. We find the overlap between the two states in next section.

\subsection{Fidelity of final state at receiver}
\label{Fidelity}
In order to find the overlap of the quantum state obtained by Bob with the actual quantum state $|\varphi \rangle = \alpha | 10\rangle +\beta | 01\rangle$, that Alice wanted to teleport, we use fidelity~($F$) as a measure, with  $0 \leq F \leq  1$. For $F=0$, there is no overlap, whereas for $F=1$, the overlap is maximum. Fidelity
 \begin{align} 
   F^{qrst}_{c_{H},c_{V}}=&\sqrt{\langle\varphi|\varrho_{c_{H},c_{V}}^{qrst}|\varphi\rangle}.
 \end{align}
is calculated using Eqs.~\eqref{eq:pureteleportedstate}~and~\eqref{eq:state}~as
 \begin{align}
   F^{qrst}_{c_{H},c_{V}}&=\textbf{[}\beta^{2}(P_{2000}^{qrst}+P_{0020}^{qrst})+\alpha^{2}(P_{0200}^{qrst}+P_{0002}^{qrst})\nonumber\\&+P_{1100}^{qrst}+P_{0011}^{qrst}+(\alpha^{2}-\beta^{2})^{2}(P_{1001}^{qrst}+P_{0110}^{qrst})\textbf{]}^{\frac{1}{2}}. 
  \end{align} 
  Using Eq.~\eqref{eq:posteriorprobability} in above equation, finally we have fidelity in the following form
  \begin{widetext}
  \begin{align}
   F^{qrst}_{c_{H},c_{V}}=&\label{finalfidelity}{\LARGE \textbf{[}} \frac{\tanh^{4}\chi}{2\times Z^{\tiny{qrst}}}\{\alpha^{2}\beta^{2}(p(qrst|2000)+p(qrst|0020)+p(qrst|0200)+p(qrst|0002))+\frac{1}{2}(p(qrst|1100)\nonumber\\&+p(qrst|0011)+(\alpha^{2}-\beta^{2})^{2}(p(qrst|1001)+p(qrst|0110)))\}{\LARGE \textbf{]}}^{\frac{1}{2}}. \nonumber\\
  \end{align}  
  \end{widetext}
Note that $p(qrst|ijkl)$ are conditional probabilities of detecting photons
by four threshold detectors given by Eq.~\eqref{eq:independentprobabilities} and value of $Z^{qrst}$ is given by Eq.~\eqref{eq:fracprobability}. Thus we have developed a closed form solution of the fidelity between the teleported state and the actual desired state. 
\begin{figure}[h]
\includegraphics[width=8.2cm,height=6.0cm]{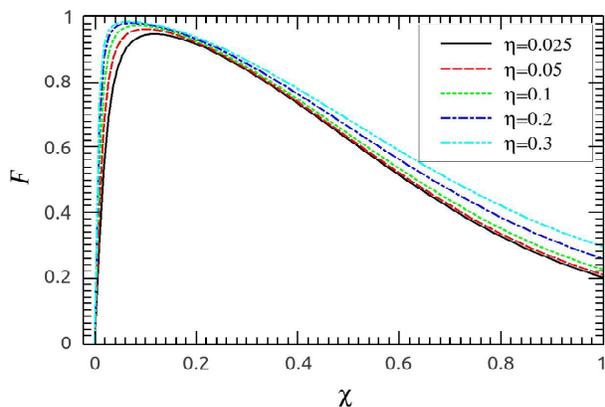}
\caption{(Color online)
Plot of fidelity $F$ against the square root of photon-pair production rate, $\chi$ for dark-count probabilities, $\zeta _{dc}=10^{-5}$, and various efficiencies $eta=0.025$ to 0.3 from the curves of lowest to highest fidelity.}
\label{fig:FidvsChidiffetas}
\end{figure}

\begin{figure}[h]
\includegraphics[width=8.2cm,height=6.0cm]{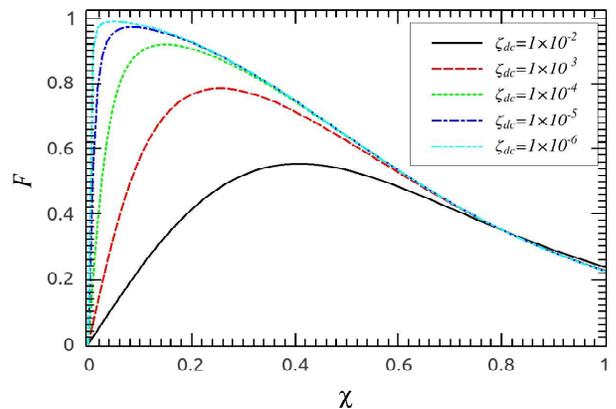}
\caption{(Color online)
Plot of fidelity $F$ against the square root of photon-pair production rate, $\chi$ for
fixed efficiency $\eta=0.1$ and dark-count probabilities, $\zeta _{dc}$,  $10^{-2}$ to $10^{-6}$ from the curves of lowest to highest fidelity.}
\label{fig:Fidelityfordifferentdarkcounts}
\end{figure}

\begin{figure}[h]
\includegraphics[width=8.2cm,height=6.0cm]{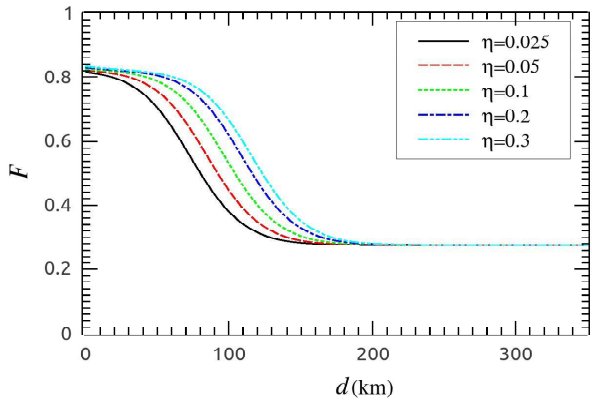}
\caption{(Color online) Fidelity $F$ is plotted vs distance $d$~km for  $\chi=0.316$ and 
dark-count probability, $\zeta _{dc}=10^{-5}$,  and $\eta=0.025$ to 0.3 (from lower to higher fidelity curves).}
\label{fig:Fidelityvsdistdiffetas}
\end{figure}

 We now check the effect of the main limiting parameters in our QT theory. We plot fidelity against square root of photon pair production rate $\chi$, for fixed dark counts probability $\zeta _{dc}=10^{-5}$  and different detectors efficiencies in Fig.~\ref{fig:FidvsChidiffetas}. From Fig.~\ref{fig:FidvsChidiffetas} it is clear that above the classical limit, fidelity does not change prominently with changing detector efficiencies. It should be noted that there is region for some small value of $\chi$ where we can achieve unit fidelity irrespective of detectors efficiencies. Variation of fidelity with $\chi$ for different dark count probabilities is given in Fig.~\ref{fig:Fidelityfordifferentdarkcounts}, which shows that high dark counts effect the fidelity dominantly. Fig.~\ref{fig:Fidelityvsdistdiffetas} shows the variation of fidelity vs distance.The fidelity saturates for certain distance but as soon as the dark counts become effective, it drops down suddenly.

We now apply our model to a recent experiment in next section.

\section{ Comparison with experimental QT}
\label{comparison}
In order to test the validity of our model of QT, we apply it to a recent experiment~\cite{YRLC+12}. In this experiment long distance~QT is achieved for over~100~km. We compare the fidelity of the teleported state with that of the experiment using the same values of resource parameters such as photon pair production rate, $\chi$, detectors efficiency, $\eta$, and dark  count rate as in the experiment. This allows us to check the validity of our model for long distance~QT and to analyze the effect of transmission losses as well. It is important to note that in our theory transmission loss is included in detection efficiency.  

\begin{figure}[h]
 \includegraphics[width=8cm,height=5.5cm]{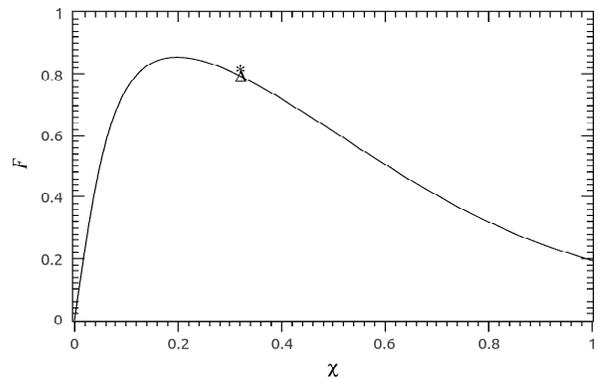}
\caption{
Fidelity F versus the square root of  photon-pair production rate, $\chi$,
 with fixed dark-count probability  $\zeta _{dc} = 1\times 10^{-6}$ and detectors efficiency value: $\eta =0.7463\times 10^{-5}$ including transmission losses. Calculated average fidelity is $79.8 \%$, at $\vartriangle$, and experimental value is $81.35\%$., at $\ast$. }\label{fig:experimentalvalues}
\end{figure}

We compare the predictions of our model with a recent experiment of QT using polarization property of photons~\cite{YRLC+12}. The conditions of this experiment are given by approximate values: detectors efficiency is $\eta\approx0.236$, transmission line has loss coefficient 45~dB, $\chi\approx \sqrt{0.1}\approx0.316$ and dark counts rate is $200s^{-1}$. Figure.~\ref{fig:experimentalvalues} demonstrates, our numerical simulation result of fidelity using these parameters. In our case the average fidelity is~$79.8\%$, whereas the average fidelity obtained in the experiment is~$81.35 \% $ with $\pm1$ standard deviation for~$|H\rangle$,~$|V\rangle$ and~$|\pm\rangle$ teleported states. 
It is interesting to observe that the photon-pair production rate of the PDC source used in this experiment, $\chi\approx 0.316$, lies far beyond its optimal value. The small difference in the calculated value by our model and the experimental value of average fidelities is due to the fact that detectors inefficiency and multi-photon pair production are not the only practical limitation that effect the fidelity. Other limitations include the imperfect entanglement generated by source, even in the scenario  where only one single pair is created. In addition multimode analysis of the setup is needed for exact modeling of the experiment. 

\begin{figure}[h]
 \includegraphics[width=8cm,height=5.5cm]{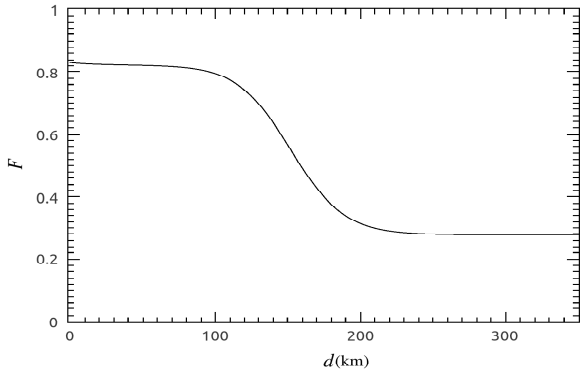}
\caption{
Fidelity $F$ vs distance $d$ 
 with dark-count probability  $\zeta _{dc} =10^{-6}$ and detectors efficiency $\eta =0.236$ at $\chi=0.315$. }\label{fig:experimentalvaluesdist}

\end{figure}

Figure~\ref{fig:experimentalvaluesdist} shows the variation of fidelity with distance for the same experimental parameters. It is apparent that with these parameters a maximum distance of around 100km can be achieved. After this distance, the dark counts become effective and fidelity decreases abruptly. The initial sub unit fidelity is due to the presence of multipairs, which lead to spurious coincidences. For detectors with unit efficiency and zero dark counts visibility will saturate to this sub unit fidelity for asymptotically large distance.

\section{Conclusion}
\label{sec:conclusions}
We have developed the model for practical quantum teleportation. We have used spontaneous parametric down conversion (SPDC)  TYPE-II source for pre-shared entangled pair between the sender and receiver. Performing Bell measurement on the unknown state and the sender part of the entangled pair, we are able to find the teleported state. We have incorporated the faulty apparatus including multipair SPDC sources, detector inefficiencies and dark counts and channel losses. We have got a closed form for the teleported state. This teleported state has led us to calculate the fidelity of the teleported state with the desired unknown quantum state in terms of nonlinearity,  $\chi^{(2)}$, of the source, detectors efficiency $\eta$ and dark counts probability $\zeta_{dc}$. 

We show the variation of fidelity with the photon pair production rate, $\chi$,  while fixing either dark counts, $\zeta_{dc}$, and varying detectors efficiency, $\eta$, or vice a versa. Our theory gives expected results of the variation of fidelity the efficiency $\eta$ and dark count probability of detectors. By increasing photon pair production rate $\chi$, fidelity first increases for low $\chi$ and after reaching a maximum, it decreases as the multipair effect becomes dominant. This dependence is due to the fact that for small value of $\chi$ the probability of photon pair generation is small and thus fidelity obtained is small because we do not have entangled photon pair due to vacuum component, but as soon as $\chi$ reaches certain value which generate entangled photon pair with very less multi pair photons generation, fidelity reaches certain maximum value which is far greater than  classical value 0.66~\cite{P94}. A very valuable conclusion we get from our investigation that the high photon pair production rate is counterproductive. If we increase the value of $\chi $, after certain value of $\chi=6\times 10^{-2}$, fidelity starts decreasing. After further increasing $\chi$ the fidelity decreases because of the events due to multipair photon generation.

   We apply our model to a recent experiment on quantum teleportation~\cite{YRLC+12}. Using same practical parameters given in ~\cite{YRLC+12}, we obtained average fidelity of $79.8\%$. Our calculated fidelity is in good agreement with the experimental value of $81.35 \% $. Thus our model fits very well to the real world experimental conditions. Our model can thus be used to predict the optimal parameters for the resources used to get the desired fidelity.  
  
\acknowledgments
We thank Barry C. Sanders for helpful comments and valuable discussions. 

\bibliographystyle{apsrev}
\bibliography{References}
\end{document}